# Can biophysics tell us something about the weak equivalence principle vis a vis the thought experiment of Einstein involving human subjects?

Fred H. Thaheld
fthaheld@directcon.net

**Abstract**

   Over a period of several decades it has been noticed that most astronauts, either orbiting the earth or on trips to the moon, have observed *phosphenes* or *light flashes* (LF) including *streaks*, *spots* and *clouds of light* when their eyes are closed or they are in a darkened cabin. Scientists suspect that two separate components of cosmic rays cause these flashes due to direct interaction with the retina. This phenomenon is not noticed on the ground because of cosmic ray interaction with the atmosphere.  The argument is advanced that this effect may provide us with a new method of exploring the weak equivalence principle from the standpoint of Einstein's original thought experiment involving human subjects.  This can be done, utilizing the retina only, as an *animate* quantum mechanical measuring device or, in conjunction with the Anomalous Long Term Effects on Astronauts (ALTEA) facility.

**Introduction**

   Is it possible that the thought experiment proposed by Einstein, regarding the weak equivalence principle (**WEP),** no longer reflects a realistic position and, may have been circumvented or replaced by virtue of recent astronaut observations of *phosphenes* or *light flashes* (LF)  in the eye or retina resulting from incident cosmic rays while in space (Casolino, 2003; Fuglesang et al, 2006; Fuglesang, 2007; Narici, 2006, 2008b)?

This also raises the question as to whether any importance can be attached to such a modification proposed herein of Einstein's original thought proposal, in light of the many sophisticated experiments which have been performed over several decades, which continue to show, with increasing orders of magnitude, the apparent inviolability of the **WEP**, utilizing *inanimate* measuring instruments (Will, 2006). It must be stressed however, that much more sophisticated satellite Eötvös tests of the **WEP** for zero-point vacuum energy will be undertaken in the near future, with an increase in baseline sensitivity to **WEP** of several orders of magnitude (Moffat, Gillies, 2002). Any observed violation of the **WEP** for vacuum energy density would constitute a significant clue as to the origin of the cosmological constant and the source of dark energy, and put the main theme of this paper on a much firmer footing.

It will be shown in this paper that repeated tests of the thought experiment proposed by Einstein appear to have been carried out in a fairly uncomplicated fashion involving the retina and cosmic rays. And, what is even more amazing, that these experiments concerning the **WEP** have been going on for years on a daily basis, quite unplanned and unknowingly, by both astronauts in space and the population on earth!

**The weak equivalence principle**

In order to properly set the stage for the proposed experiments, and an analysis of what appears to have been transpiring unbeknownst to us, it is helpful to briefly and simply summarize the **WEP** in the following statements (Dunsby, 1996):

**WEP 1**: There are no local experiments which can distinguish non-rotating free fall in a gravitational field from uniform motion in space in the absence of a gravitational field. By local we mean that observations are confined to a region over which the variation of the gravitational field is un-observably small.

**WEP 2:** A frame linearly accelerated relative to an inertial frame in special relativity is locally identical to a frame at rest in a gravitational field. Let us examine both of these statements from the standpoint of a human observer, just as Einstein originally outlined.

As regards **WEP 1**, an observer is in a rocket ship with no windows in it, or "other methods of communication with the outside world", undergoing uniform motion in a part of the universe far removed from gravitating bodies. A released body is found to remain at rest relative to the observer in the rocket ship.

He is next placed in a lift in an evacuated shaft and is allowed to fall freely towards the center of the earth. A released body is found to remain at rest relative to the observer. From the point of view of the observer in the lift, both cases are indistinguishable. No measuring instruments that operate completely inside the lift are able to distinguish between the two cases.

As regards **WEP 2**, he is again placed in a rocket ship in a part of the universe far removed from any gravitating bodies. The rocket ship is accelerated forward with a constant acceleration $g$ relative to an inertial observer. The observer releases a body from rest and sees it fall to the floor with acceleration $g$. The rocket ship is next placed on the surface of the earth, whose rotational and orbital motions are ignored. A released body is found to fall to the floor with acceleration $g$. Once again, both cases are indistinguishable and no measuring instruments operating completely inside the rocket ship can distinguish between them. Will this sequence of events still be found to be true if we examine this from the standpoint of recent observations made by astronauts of LF?

**Astronaut observations of phosphenes from cosmic rays**

By way of a brief introduction, cosmic rays are high energy charged particles originating in outer space, that travel at nearly the speed of light, and whose existence in space is *isotropic* (Mewaldt, 1996; Semyonov, 2006). Most cosmic rays are the nuclei of atoms, ranging from the lightest to the heaviest elements in the periodic table. They also include high energy electrons, positrons and other subatomic particles such as muons and pions. Approximately 89% of the nuclei are H (protons), 10% He and 1% heavier elements.

Ever since the Apollo missions, astronauts have reported seeing LF in space, usually when they are in a darkened cabin or when they close their eyes (Casolino, 2003; Fuglesang et al, 2006, 2007; Narici et al, 2006, 2008a; Nurzia et al, 2005; Pinskey et al, 1974, 1975). These LF in space were even predicted before the first space mission and possible radiation hazards were pointed out (Tobias, 1952) . Scientists now suspect that two separate components of cosmic rays cause these flashes in a complementary fashion, one due to heavy cosmic particles such as He and Li, and one due to lighter H protons (Nurzia et al, 2005). They hypothesize that the direct interaction of the heavy nuclei with the retina, causes ionization or excitation and, in addition, that the proton-induced nuclear interactions in the eye (with a lower interaction probability) produce knock-on particles. This is a variable process depending upon the geomagnetic cutoff above the earth, average orbital heights and space station shielding and, varies among individual astronauts as to intensity of, and time between, perception events (Fuglesang et al, 2006).

Cosmic rays are found in great abundance throughout the universe, with a wide range of particle varieties and energies, far from any gravitational influences. And, no special instrument is needed to detect their presence in this instance, only a human subject. It can be said that when these particles impinge upon the retina, this constitutes a *measurement*, and it is this *objective* information which is conveyed to the visual cortex of the human (Thaheld, 2005a, 2005b, 2008). It is also of interest to note here that this phenomenon provides the only way to really see an elementary particle, leading one to speculate that a quantum microscopic particle can directly tell us something about a macroscopic gravitational state in an *animate* setting, and without the need for any instruments (Nurzia et al, 2005).

This LF effect is never seen by humans on the surface of the earth nor by passengers on airliners or military aircraft, due to the interaction between the earth's atmosphere and the heavy cosmic particles and the protons, which interaction results in much lower particle energy showers, which have no discernible effect on the retina at lower altitudes. The exception to this is when patients are subjected to ion therapy for brain tumors (Narici, 2008a). It is essential for these proposed experiments that the cosmic rays are *isotropic* when we proceed further into space away from the earth. And, that the effects of any neutral or ionized interstellar gas can be ignored if one remains below relativistic velocities (Semyonov, 2006).

**Proposed experiment with the eyes alone**

I am sure you can all see where this is proceeding. As regards **WEP 1**, if it was possible for one to be in a lift in freefall towards the center of the earth, and closes his eyes or reduces the ambient light, he will see no LF, and thereby knows that he is not out in space undergoing uniform motion. If however, he closes his eyes or reduces the light, and sees these LF, he will know that he is in outer space undergoing uniform motion. In both instances he will know what is causing the action on the released floating object i.e., quantum mechanics is telling him something about relativity and gravity! The only *measuring instrument* that is required in this instance is the eye or retina, in conjunction with the brain and the visual cortex.

It is obvious that the same argument can be applied to **WEP 2**, in that if one is in a rocket ship on the surface of the earth and closes his eyes or reduces the ambient light, he will see no LF, and will thereby know that he is not out in space undergoing acceleration. If however, he closes his eyes or reduces the light, and sees these LF, he will know that he is in outer space undergoing acceleration. Again, he can differentiate between the effects of gravity and acceleration *g* on a released body through an *animate* quantum mechanical process. And, once again the only *measuring instrument* is the eye, and there is no need for any other type of instrumentation for detection purposes.

**Experiment with the eyes and the ALTEA facility**

ALTEA consists of a silicon detector system (SDS) positioned around the astronaut's head on a helmet shaped holder containing a 32 electrode EEG cap, including 3 floating electrodes for retinogram (ERG) measurements (Narici, 2006, 2008a, 2008b). It also has a visual stimulator unit (VSU) which delivers the light stimulation paradigm for the visual evoked potential (VEP). The SDS is able to reconstruct the trajectories of specific cosmic ray particles and ascertain what type, while correlating this with electrophysiological readings such as ERG, VEP and EEG. ALTEA measures the particles passing through the astronaut's eyes/brain, their electrophysiological brain dynamics and the visual system status, and each perception of a LF is signaled with a pushbutton. Objective brain signals or EEGs are correlated with subjective LF as well as particles.

When the astronaut is utilizing the ALTEA facility, he is able to perform all the same **WEP** experiments as before, only with greater accuracy, as all the essential ingredients leading up to a LF, such as the type of cosmic ray particle, the energy, trajectory and number, are being recorded in real time along with the ERG, EEG and VEP readings vs the earlier somewhat crude anecdotal responses regarding just the LF. The importance of ALTEA will be explored in the Discussion section.

It has been shown that when ALTEA was operated in the dosimetry (DOSI) mode, it revealed a detailed spectrum of the radiation environment, showing some 23 ions ranging from B to Fe (Narici, 2006, 2008a).

**Experiment utilizing retinal tissue on microelectrode arrays**

It may also be possible to simultaneously conduct corroborating backup experiments using excised retinal tissue mounted on microelectrode arrays (MEAs), in a similar fashion as has been previously proposed (Thaheld, 2003). Over the years a considerable amount of research has been performed on complete pieces of retina obtained by cutting the eyecup into segments (Stett et al, 2000). Certain outer elements such as the sclera, pigment epithelium and the inner limiting membrane, which normally account for large photon transmission losses, are then removed. These pieces are then mounted on a planar array consisting of > 60 microelectrodes which are capable of recording the individual extracellular action potentials from > 100 ganglion cells (Meister et al, 1994).

This retinal tissue can be directly exposed to collimated ion beams from accelerators in much the same manner as the mouse retina was to $^{12}$C ions, taking care not to activate any of the microelectrodes with the ions and corrupt the resulting data (Sannita et al, 2007). This procedure can be undertaken prior to putting this retinal tissue with the MEAs into space.

Since this retinal tissue will be directly exposed to any incident cosmic rays vs the retinal tissue which is normally shielded by the brain, and related outer tissues which have been removed, we may observe more cosmic ray initiated events than are visualized or reported on by the astronaut. This technique may allow one to obtain more accurate readings for statistical purposes, of an electrically amplified nature, which would serve to complement the anecdotal responses coming from the astronaut coupled with the ALTEA results, as to the exact number and intensity of these LF. The major drawback to such a proposal at the present time is the limited life span of the retinal tissue, especially if its use is contemplated in space.

**The relationship between phosphenes, ERGs and the measurement problem**

Could these LF have a bearing on a resolution of the measurement problem? A system is characterized by state vectors or wave functions which change in two ways, continuously in a linear fashion over time and discontinuously if a measurement is made (Thaheld, 2005a; 2008). The second process is referred to as the collapse of the wave function, and the measurement problem arises as to how and when the wave function collapses or how a state reduction to one of the eigenstates of the measured observable occurs. If appears that these LF may provide support for a theory advanced by the author that wave function collapse takes place in the rhodopsin or retinal molecule (Thaheld, 2005a; 2008; 2009).

The ERG records the electrical response of the retina to photic and cosmic ray stimulation, and has also been used with mouse retina irradiated with $^{12}$C ions from an accelerator (Sannita et al, 2007). To directly quote, "The problem is that despite the similarities with the waveform of light-evoked ERG, the origin of the retinal response to $^{12}$C ions remains to be defined. The intensities of photons and ionizing radiation are not comparable and comparison between their effects should therefore be cautious, with due concern for the stimuli differences". There may be a way to examine this issue by recording a component of the ERG known as the early receptor potential (ERP), (Sakmar, 1999). The ERP is a biphasic response comprising an initial cornea-positive fast phase followed by a slower cornea-negative phase. The action spectra matches that of rhodopsin, and the amplitudes of both phases are linearly proportional to the fraction of rhodopsin bleached by a stimulus flash. The ERP reflects directly dipole changes in the visual pigment molecules due to conformational changes that are elicited by photon absorption. I.e., the net displacement of electric charge in rhodopsin molecules generates the ERP.

The problem is that while photons and $^{12}$C ions can both cause a conformational change of the rhodopsin molecule from *cis* to *trans*, we know that the photons achieve this by being absorbed by the molecule in a π-π* electron orbital transition, while the more energetic $^{12}$C ions may accomplish this in a different fashion (Thaheld, 2008). However, the end result is the same, in that the wave function is collapsed (Thaheld,

2005a).

**Discussion**

An objection might be raised to this proposed experiment regarding the fact that cosmic rays can be considered as "other methods of *communication* with the outside world", although this caveat has never been raised before. Indeed, just such an objection has only recently been raised (Parker, 2006) in the following manner: "In reading through your paper, it seems to me that the cosmic rays responsible for the scintillations in the eyes are equivalent to a different form of window to the outside world. Instead of looking out through a window to see if you are in space or sitting on the surface of a planet, etc. the cosmic rays penetrate through the walls of the laboratory to tell you the same thing. If you built a truly isolated laboratory, with massive walls that stop all cosmic rays, you would be back to the basic principle of equivalence". The problem with this approach is that a truly isolated laboratory in space with massive walls (you would not need this arrangement on the surface of the earth due to the atmosphere), would require shielding weighing tens of tons (Parker, 2006).

Another objection has been raised that this paper has essentially nothing to do with the **WEP**, beyond a quibble over semantics. That an observer can look out a window or use radar or a gamma ray detector. Except, that none of these options would elicit a quantum mechanical response from the eye or retina in the form of anomalous LF.

Setting aside these objections for the purpose of discussion, it would appear that we have already achieved meaningful results, since we can combine existing observations as they might relate to both **WEP 1** and **WEP 2** in an unusual fashion. We already know over a period of years and on a continuing daily basis, that human subjects in space undergoing uniform motion while in orbit, repeatedly observe these LF, which directly relates to half of the pronouncement of **WEP 1**. We also know, again over a period of years and on a continuing daily basis, that human subjects on the surface of the earth never see any LF, which directly relates to half of the pronouncement of **WEP 2**. One can then surmise, with great certainty, that the remaining pronouncements of both **WEP 1** and **WEP 2**, concerning free fall towards the center of the earth (unattainable) within the framework of **WEP 1,** and constant acceleration *g* in outer space as per **WEP 2**, would be found to be similarly applicable, even without confirming experiments.

One of the most interesting aspects that comes to mind, has to do with the matter of time and relativity, since an astronaut will immediately know that he is accelerating at *g* due to the LF in his retina and the information provided by ALTEA and the MEAs. This will put him in a unique position as regards the evaluation of time, from both a quantum mechanical and relativistic standpoint. When he looks at a clock on board his rocket ship, he will observe it as being normal, even though it is running slower to an outside observer in an inertial reference frame, due to relativistic effects arising from an increase in velocity. However, the flashes of light due to quantum mechanical effects, *simultaneously* informs the astronaut that the clock is running slow, although this information may be of an inexact and gross nature, if he is not simultaneously observing the input from ALTEA and the retinal tissue on the MEAs. So, he is cognizant of both inertial and non-inertial frames, although it will be of an *all* or *none* nature if he is only relying upon his eyes and the resulting LF. He may not know exactly how much slower his clock is running, like an outside observer in an inertial frame would but, he *does know* that it is running slower.  Also, he may not be able to determine from an increase in flashes in his retina what his velocity really is, whether of a nonrelativistic or relativistic nature.

Since the cosmic rays are already traveling at close to the speed of light, for a long period during the acceleration *g* of the spacecraft, we will have a situation where the energies of these particles impinging essentially head-on, will be increasing relative to the retina, while those cosmic rays approaching essentially from the rear will correspondingly have their energies reduced (Semyonov, 2006).  At this point in time we do not know the minimal cosmic ray energies which can still cause this phenomenon but, the fact that they are already moving at close to the speed of light, probably means that these particles approaching from the rear will still possess sufficient energy over a wide range of velocities (into the relativistic) to still cause this phenomenon.  We also do not know what effect the particles will have that are impinging with increased velocities and energies upon the retina, although one can hazard a guess that either the number of flashes and/or their intensity will increase, while for those approaching from the rear they will decrease.

How could one accurately determine at what velocity they were proceeding and how slow their clock was really running? This would appear difficult to achieve when one considers that present astronaut observations of LF vary over a wide range of time, with some reporting no LF at all (Fuglesang, 2007).  One also has to contend with a wide range of particles and velocities, impinging from all directions and perhaps not truly *isotropic* over the short term (Semyonov, 2006).

One approach may be to take readings while ALTEA is facing in the direction of acceleration, moving into the cosmic ray flux, and then rotating ALTEA so it is facing in the opposite direction, moving away from the cosmic rays. The total data derived from all measurements should allow one to roughly determine the velocity at which they are proceeding, and from that the quantum mechanical clock adjustment. The solution to this problem is really not that simple but, will hopefully lead to other proposals involving either measuring the total cosmic ray flux or shielding out specific particles. For a hint as to other solutions, involving possible shielding options, I would recommend the excellent paper by Semyonov, exploring the radiation hazards of relativistic interstellar flight (Semyonov, 2006).

These unknown effects, when coupled with the results obtained from the retinal tissue on the MEAs, may enable the astronaut to determine much more precisely at what velocity he is proceeding, whether of a non-relativistic or relativistic nature. I.e., the increased energy of the heavy nuclei may cause considerably more ionization or excitation of the retina and, the proton-induced nuclear interactions in the eye will produce considerably more knock-on particles.

The question now before us is as follows: Will these proposed experiments have any bearing upon the validity of the **WEP** and its relationship to special and general relativity and, quantum mechanics, in the light of such previously mentioned experiments exhaustively confirming the principle of equivalence? I.e., is this effect of cosmic rays just a variation or additional caveat which should be made to the existing Einstein thought experiments and, one should not fall into that seductive trap of conveniently reading more into one's own thought experiments than really exists in the first place?

The following comment has been made upon this **WEP** approach (Matsuno, 2006). "Your paper just reminds me of a possible relationship between **WEP** and quantum mechanics. Once we accept both **WEP** and the invariance of light velocity, general relativity remains intact. If we try to save some room for quantum mechanics on the other hand, we would also have to do something with **WEP** since the invariance of light velocity seems to remain incontestable. Your quote of astronaut's flashing experiences must be an instance of measurement. This must be the case that QM is quintessential. The relationship between **WEP** and the objective reduction of the wave function seems to be a no-man's land as of now".

There is one other issue which should be further addressed, and it revolves around the matter of the hazards which these cosmic rays pose for the astronauts (Sannita et al, 2006, 2007). The astronaut out in space faces a big problem. Whether he is being *accelerated* forward with a constant acceleration *g* (even if he remains below relativistic velocities) or being subjected to uniform motion, cosmic rays will be constantly impinging upon his retina, brain and body over a long period of time, with potentially very harmful effects (Parker, 2006). In fact, it has been noticed that a high % of astronauts develop cataracts in their eyes in later years.

**Conclusion**

1. Astronauts out in space have been observing *phosphenes* or anomalous LF in their retina and occipital cortex caused by cosmic rays. That it may be possible to utilize this phenomenon to explore Einstein's **WEP** in an *animate* quantum mechanical fashion based upon the isotropic nature of these cosmic rays.

2. That an astronaut may be able to *simultaneously* experience two different frames, one *inertial* and the other *non-inertial*, while undergoing a constant acceleration *g*, and, that he will be able to determine both velocity and clock times, varying from non-relativistic to relativistic, by utilizing additional information derived from ALTEA and the MEAs.

*3.* That the *phosphenes* may have a direct bearing on a resolution of the *measurement problem*, helping us to rule out several conflicting interpretations (Narici et al, 2009; Thaheld, 2005a; 2008; 2009).

4. If any of the proposed satellite Eötvös tests of the **WEP** for zero-point vacuum energy are successful, this may enable us to conduct biological Casimir-type experiments to determine if there is some type of coupling to the vacuum energy for a biological system (Pizzi et al, 2004, 2007; Wetz, 2001; 2002).

Finally in closing, and even as this is being written, this unusual and unheralded **WEP** experiment involving cosmic rays and LF, is taking place on the orbiting space station among a few astronauts, while the non-flashing of lights is occurring among the billions of people on earth on a daily basis.

**Acknowledgements**

I wish to thank the editor and the reviewers for their helpful comments which enabled me to broaden the overall approach taken in this paper and to clarify certain issues.